\documentclass[conference]{IEEEtran}

\renewcommand\IEEEkeywordsname{Index Terms}
\usepackage{graphicx}
\usepackage{subcaption}
\usepackage{tikz,xparse}

\usetikzlibrary{patterns}

\usetikzlibrary{dsp,chains}
\usetikzlibrary{matrix}
\usepackage{mathptmx}
\DeclareMathAlphabet{\mathcal}{OMS}{cmsy}{m}{n}
\usepackage{verbatim}
\usepackage{calc}
\usepackage{ifthen}
\usepackage{xifthen}
\usepackage{cancel}
\usepackage{bm}
\usepackage{verbatim}
\usepackage{multirow}
\usepackage{cite}
\usepackage[hyphens]{url}
\usepackage[nolist]{acronym} 
\usepackage{pgfplots}
\usetikzlibrary{arrows,shapes,graphs,graphs.standard,quotes,arrows.meta,decorations.markings}
\pgfplotsset{compat=newest}
\usepackage[bookmarks=false]{hyperref}
\usepackage{units}
\usepackage{amsmath, amsbsy, amssymb, latexsym }
\hypersetup{bookmarksdepth=-2}
\usepackage{comment}
\usepackage[utf8]{inputenc}
\usepackage{xcolor}
\usepackage{enumitem}

\usepackage{marginnote}
\tikzset{>=latex}

\definecolor{mittelblau}{RGB}{0, 126, 198}
\definecolor{violettblau}{cmyk}{0.9, 0.6, 0, 0}
\definecolor{rot}{RGB}{238, 28 35}
\definecolor{apfelgruen}{RGB}{140, 198, 62}
\definecolor{gelb}{RGB}{255, 229, 0}
\definecolor{orange}{RGB}{244, 111, 33}
\definecolor{pink}{RGB}{237, 0, 140}
\definecolor{lila}{RGB}{128, 10, 145}
\definecolor{hellgrau}{RGB}{224, 224, 224}
\definecolor{mittelgrau}{RGB}{128, 128, 128}
\definecolor{dunkelgrau}{RGB}{80,80,80}
\definecolor{anthrazit}{RGB}{19, 31, 31}
\definecolor{darkgreen}{RGB}{34,139,34}
\definecolor{aqua}{RGB}{0, 255, 255}
\tikzset{
       vnd/.style={
        shape=circle,
        fill=black,
        draw,
        inner sep=0pt,
        minimum size=0.2cm},
        cnd/.style={
        shape=rectangle,
        fill=white,
        draw,
        minimum width=0.05mm,
        minimum height = 0.05mm}, 
         vndR/.style={
        shape=circle,
        fill=red,
        draw,
        inner sep=0pt,
        minimum size=0.2cm},
        cndR/.style={
        shape=rectangle,
        fill=white,
        draw=red,
        minimum width=0.05mm,
        minimum height = 0.05mm}
}

\IEEEoverridecommandlockouts

\renewcommand{\vec}[1]{\mathbf{#1}}

\newcommand{\Lm}{\vec{L}}

\newcommand{\Nm}{\vec{N}}

\begin{document}

\begin{NoHyper}
\title{A Polar Subcode Approach to Belief Propagation List Decoding}

\author{\IEEEauthorblockN{Marvin Geiselhart, Ahmed Elkelesh, Jannis Clausius and Stephan ten Brink }
	\IEEEauthorblockA{
		Institute of Telecommunications, Pfaffenwaldring 47, University of  Stuttgart, 70569 Stuttgart, Germany 
		\\\{geiselhart,elkelesh,clausius,tenbrink\}@inue.uni-stuttgart.de\\
	}
	\thanks{This work is supported by the German Federal Ministry of Education and Research (BMBF) within the project Open6GHub (grant no. 16KISK019).
}
\vspace{-0.7cm}
}

\maketitle

\begin{acronym}
\acro{ML}{maximum likelihood}
\acro{BP}{belief propagation}
\acro{BPL}{belief propagation list}
\acro{LDPC}{low-density parity-check}
\acro{BER}{bit error rate}
\acro{SNR}{signal-to-noise-ratio}
\acro{BPSK}{binary phase shift keying}
\acro{AWGN}{additive white Gaussian noise}
\acro{LLR}{log-likelihood ratio}
\acro{MAP}{maximum a posteriori}
\acro{BLER}{block error rate}
\acro{SCL}{successive cancellation list}
\acro{SC}{successive cancellation}
\acro{BI-DMC}{Binary Input Discrete Memoryless Channel}
\acro{CRC}{cyclic redundancy check}
\acro{CA-SCL}{CRC-aided SCL}
\acro{CA-BPL}{CRC-aided belief propagation list}
\acro{BEC}{Binary Erasure Channel}
\acro{BSC}{Binary Symmetric Channel}
\acro{BCH}{Bose-Chaudhuri-Hocquenghem}
\acro{RM}{Reed--Muller}
\acro{RS}{Reed-Solomon}
\acro{SISO}{soft-in/soft-out}
\acro{3GPP}{3rd Generation Partnership Project }
\acro{5G}{5th generation mobile communications}
\acro{eMBB}{enhanced Mobile Broadband}
\acro{CN}{check node}
\acro{VN}{variable node}
\acro{GenAlg}{Genetic Algorithm}
\acro{CSI}{Channel State Information}
\acro{OSD}{ordered statistic decoding}
\acro{MWPC-BP}{minimum-weight parity-check BP}
\acro{FFG}{Forney-style factor graph}
\acro{MBBP}{multiple-bases belief propagation}
\acro{URLLC}{ultra-reliable low-latency communications}
\acro{DMC}{discrete memoryless channel}
\acro{SGD}{stochastic gradient descent}
\acro{QC}{quasi-cyclic}
\acro{PR-CA-BP}{permuted relaxed CRC-aided BP}
\acro{R-CA-BP}{relaxed CRC-aided BP}
\acro{R-CA-BPL}{relaxed CRC-aided BPL}
\acro{PR-CA-BPL}{permuted relaxed CRC-aided BPL}
\acro{P-CA-BPL}{permuted CRC-aided BPL}
\acro{P-CA-BP}{permuted CRC-aided BP}
\acro{LTA}{lower triangular affine}
\acro{GA}{general affine}
\acro{AED}{automorphism ensemble decoding}
\acro{SPA}{sum-product algorithm}
\acro{PAC}{polarization adjusted convolutional}
\end{acronym}

\begin{abstract}
Permutation decoding gained recent interest as it can exploit the symmetries of a code in a parallel fashion. Moreover, it has been shown that by viewing permuted polar codes as polar subcodes, the set of usable permutations in permutation decoding can be increased. We extend this idea to pre-transformed polar codes, such as \ac{CRC}-aided polar codes, which previously could not be decoded using permutations due to their lack of automorphisms. Using \ac{BP}-based subdecoders, we showcase a performance close to \ac{CA-SCL} decoding. The proposed algorithm outperforms the previously best performing \textit{iterative} \ac{CA-BPL} decoder both in error-rate performance and decoding latency.
\end{abstract}
\acresetall



\acresetall

\section{Introduction}
Polar codes were first introduced by Ar\i kan as a class of codes that achieve the capacity of binary memoryless channels \cite{ArikanMain} using the \ac{SC} decoding algorithm. In the short block length regime, the performance of stand-alone polar codes (under optimal \ac{ML} decoding) is however not favorable. It could however be shown that a high-rate outer code, such as a \ac{CRC}, significantly improves the error-correcting performance of polar codes under \ac{SCL} decoding \cite{talvardyList}. As a consequence, \ac{CRC}-concatenated polar codes have been selected for the control channel in the \ac{5G} standard \cite{polar5G2018}.
Despite its good error-correcting performance, \ac{CA-SCL} decoding is inherently sequential and hard-output. Consequently, the decoding latency is comparably high and the algorithm is not well-suited for iterative receivers (e.g., iterative detection and decoding loops).

An alternative to \ac{SC} decoding is the iterative \ac{BP} algorithm \cite{ArikanBP,ArikanBP_original}, which is parallel by nature and, thus, better suited for high-throughput/low-latency applications. Unfortunately, its error-rate performance is generally worse than that of \ac{CA-SCL} decoding.

Over the last couple of years, numerous improvements to \ac{BP} decoding of polar codes have been proposed. Notable is the decoding over multiple permuted factor graphs \cite{Urbanke_chCsC_BP}, resulting in variations of the \ac{BP} algorithm (\ac{BPL} \cite{elkelesh2018belief}) and \ac{SC} decoding \cite{PermutedSCL}. A version of \ac{BP} decoding with an outer \ac{LDPC} code has been proposed \cite{BP_Siegel_Concatenating}, where the Tanner graph of the outer code has been attached to the left side of the polar factor graph. We note that, unlike in \ac{SCL} decoding, it turned out to be very difficult to effectively use the error-correcting capability of an outer \ac{CRC} code in iterative decoding, as its Tanner graph contains many short cycles. Therefore, all known iterative decoding schemes of CRC-aided polar codes require additional steps to reach reasonable performance.
In \cite{doan2018neural}, the edges are weighted to mitigate the effect of short cycles in the \ac{CRC} Tanner graph. In \cite{CRC_BPL_ISIT20}, it has been proposed to reduce the number of edges in the factor graph and use an ensemble of stage-permuted polar \ac{BP} decoders. 
Lastly, the method of ``polar relaxation'' has been proposed to reduce the number of cycles in the combined factor graph \cite{relaxedcrc}. 

At the same time, there have been advances in automorphism-based decoding, so-called \ac{AED}. In \ac{AED}, multiple permuted versions of the received sequence are decoded in parallel and the most promising candidate is chosen as the final codeword estimate. The used permutations come from the automorphism group of the code. \ac{AED} has been successfully applied to codes with rich symmetries, such as cyclic codes \cite{Hehn_MBBP_cyclic,Chen_Cyclic_LDPC_AED}, quasi-cyclic codes \cite{QCAED}, \ac{RM} codes \cite{rm_automorphism_ensemble_decoding} and stand-alone polar codes (i.e., polar codes without the \ac{CRC}-aid) \cite{PolarAutomorphisms_ISIT21}.

Recently, in \cite{KamenevPermutedSCwithAutomorphisms}, it has been shown that for polar codes, additional permutations (outside the code's automorphism group) can be used in ensemble decoding. The key idea is to observe that such permutations result in a polar subcode~\cite{polarsubcodes}.

In this work, we extend the polar subcode approach to iterative decoding. The main contributions of this paper are:
\begin{itemize}
    \item We generalize the work of \cite{KamenevPermutedSCwithAutomorphisms} to start with arbitrary polar subcodes, i.e., also \ac{CRC}-aided polar codes.
    \item We propose to apply the polar subcode permutation scheme to \ac{BPL} decoding of polar-\ac{CRC} codes and report both a better error-rate performance and a decreased latency compared to stage-permutation based \ac{CA-BPL} decoding \cite{CRC_BPL_ISIT20}.
    \item We show that a combination of this ensemble decoding scheme with \textit{polar relaxation} further enhances the performance without any additional overhead.
\end{itemize}

\section{Preliminaries}\label{sec:preliminaries}

\begin{figure}[t]
\begin{subfigure}[b]{0.48\linewidth}
\resizebox{1\columnwidth}{!}{
\begin{tikzpicture}[y=1cm]
\tikzstyle{every node}=[font=\Huge]
\tikzstyle{frozennode} = [dspnodefull,minimum size=2mm,rot]
\tikzstyle{normalnode} = [dspnodefull,minimum size=1mm]
\tikzstyle{normalline} = [line width = 0.5mm]
\tikzstyle{arrow} = [dspconn,normalline,-triangle 45]

\fill [white] (0,-.35) rectangle (7,7.35);

\node[frozennode] (u0) at (0.00, 7) {};
\node[frozennode] (u1) at (0.00, 6) {};
\node[frozennode] (u2) at (0.00, 5) {};
\node[normalnode] (u3) at (0.00, 4) {};
\node[normalnode] (u4) at (0.00, 3) {};
\node[normalnode] (u5) at (0.00, 2) {};
\node[normalnode] (u6) at (0.00, 1) {};
\node[normalnode] (u7) at (0.00, 0) {};
\node[dspadder] (node0) at (1.00, 7) {};
\node[normalnode] (node1) at (1.00, 6) {};
\draw[normalline] (u0)--(node0);
\draw[normalline] (u1)--(node1);
\draw[normalline] (node1)--(node0);
\node[dspadder] (node2) at (1.00, 5) {};
\node[normalnode] (node3) at (1.00, 4) {};
\draw[normalline] (u2)--(node2);
\draw[normalline] (u3)--(node3);
\draw[normalline] (node3)--(node2);
\node[dspadder] (node4) at (1.00, 3) {};
\node[normalnode] (node5) at (1.00, 2) {};
\draw[normalline] (u4)--(node4);
\draw[normalline] (u5)--(node5);
\draw[normalline] (node5)--(node4);
\node[dspadder] (node6) at (1.00, 1) {};
\node[normalnode] (node7) at (1.00, 0) {};
\draw[normalline] (u6)--(node6);
\draw[normalline] (u7)--(node7);
\draw[normalline] (node7)--(node6);
\node[dspadder] (node8) at (2.50, 7) {};
\node[normalnode] (node9) at (2.50, 5) {};
\draw[normalline] (node0)--(node8);
\draw[normalline] (node2)--(node9);
\draw[normalline] (node9)--(node8);
\node[dspadder] (node10) at (3.00, 6) {};
\node[normalnode] (node11) at (3.00, 4) {};
\draw[normalline] (node1)--(node10);
\draw[normalline] (node3)--(node11);
\draw[normalline] (node11)--(node10);
\node[dspadder] (node12) at (2.50, 3) {};
\node[normalnode] (node13) at (2.50, 1) {};
\draw[normalline] (node4)--(node12);
\draw[normalline] (node6)--(node13);
\draw[normalline] (node13)--(node12);
\node[dspadder] (node14) at (3.00, 2) {};
\node[normalnode] (node15) at (3.00, 0) {};
\draw[normalline] (node5)--(node14);
\draw[normalline] (node7)--(node15);
\draw[normalline] (node15)--(node14);
\node[dspadder] (node16) at (4.50, 7) {};
\node[normalnode] (node17) at (4.50, 3) {};
\draw[normalline] (node8)--(node16);
\draw[normalline] (node12)--(node17);
\draw[normalline] (node17)--(node16);
\node[dspadder] (node18) at (5.00, 6) {};
\node[normalnode] (node19) at (5.00, 2) {};
\draw[normalline] (node10)--(node18);
\draw[normalline] (node14)--(node19);
\draw[normalline] (node19)--(node18);
\node[dspadder] (node20) at (5.50, 5) {};
\node[normalnode] (node21) at (5.50, 1) {};
\draw[normalline] (node9)--(node20);
\draw[normalline] (node13)--(node21);
\draw[normalline] (node21)--(node20);
\node[dspadder] (node22) at (6.00, 4) {};
\node[normalnode] (node23) at (6.00, 0) {};
\draw[normalline] (node11)--(node22);
\draw[normalline] (node15)--(node23);
\draw[normalline] (node23)--(node22);
\node[normalnode] (x0) at (7.00, 7) {};
\draw[normalline] (node16)--(x0);
\node[normalnode] (x1) at (7.00, 6) {};
\draw[normalline] (node18)--(x1);
\node[normalnode] (x2) at (7.00, 5) {};
\draw[normalline] (node20)--(x2);
\node[normalnode] (x3) at (7.00, 4) {};
\draw[normalline] (node22)--(x3);
\node[normalnode] (x4) at (7.00, 3) {};
\draw[normalline] (node17)--(x4);
\node[normalnode] (x5) at (7.00, 2) {};
\draw[normalline] (node19)--(x5);
\node[normalnode] (x6) at (7.00, 1) {};
\draw[normalline] (node21)--(x6);
\node[normalnode] (x7) at (7.00, 0) {};
\draw[normalline] (node23)--(x7);

\end{tikzpicture}
} 
\caption{\footnotesize Regular factor graph} 
\label{fig:fg}
\end{subfigure}
\hfill     
\begin{subfigure}[b]{0.48\linewidth}
\resizebox{1\columnwidth}{!}{
\begin{tikzpicture}[y=1cm]
\tikzstyle{every node}=[font=\Huge]
\tikzstyle{frozennode} = [dspnodefull,minimum size=2mm,rot]
\tikzstyle{normalnode} = [dspnodefull,minimum size=1mm]
\tikzstyle{normalline} = [line width = 0.5mm]
\tikzstyle{arrow} = [dspconn,normalline,-triangle 45]

\fill [white] (0,-.35) rectangle (7,7.35);

\node[frozennode] (u0) at (0.00, 7) {};
\node[frozennode] (u1) at (0.00, 6) {};
\node[frozennode] (u2) at (0.00, 5) {};
\node[normalnode] (u3) at (0.00, 4) {};
\node[normalnode] (u4) at (0.00, 3) {};
\node[normalnode] (u5) at (0.00, 2) {};
\node[normalnode] (u6) at (0.00, 1) {};
\node[normalnode] (u7) at (0.00, 0) {};

\draw[style=dashed, line width = .5mm, black] (0.7,5.7) rectangle (1.3,7.3);
\node[dspadder] (node2) at (1.00, 5) {};
\node[normalnode] (node3) at (1.00, 4) {};
\draw[normalline] (u2)--(node2);
\draw[normalline] (u3)--(node3);
\draw[normalline] (node3)--(node2);
\draw[style=dashed, line width = .5mm, black] (0.7,-0.3) rectangle (3.3,3.3);
\node[dspadder] (node8) at (2.50, 7) {};
\node[normalnode] (node9) at (2.50, 5) {};
\draw[normalline] (u0)--(node8);
\draw[normalline] (node2)--(node9);
\draw[normalline] (node9)--(node8);
\node[dspadder] (node10) at (3.00, 6) {};
\node[normalnode] (node11) at (3.00, 4) {};
\draw[normalline] (u1)--(node10);
\draw[normalline] (node3)--(node11);
\draw[normalline] (node11)--(node10);
\node[dspadder] (node16) at (4.50, 7) {};
\node[normalnode] (node17) at (4.50, 3) {};
\draw[normalline] (node8)--(node16);
\draw[normalline] (u4)--(node17);
\draw[normalline] (node17)--(node16);
\node[dspadder] (node18) at (5.00, 6) {};
\node[normalnode] (node19) at (5.00, 2) {};
\draw[normalline] (node10)--(node18);
\draw[normalline] (u5)--(node19);
\draw[normalline] (node19)--(node18);
\node[dspadder] (node20) at (5.50, 5) {};
\node[normalnode] (node21) at (5.50, 1) {};
\draw[normalline] (node9)--(node20);
\draw[normalline] (u6)--(node21);
\draw[normalline] (node21)--(node20);
\node[dspadder] (node22) at (6.00, 4) {};
\node[normalnode] (node23) at (6.00, 0) {};
\draw[normalline] (node11)--(node22);
\draw[normalline] (u7)--(node23);
\draw[normalline] (node23)--(node22);
\node[normalnode] (x0) at (7.00, 7) {};
\draw[normalline] (node16)--(x0);
\node[normalnode] (x1) at (7.00, 6) {};
\draw[normalline] (node18)--(x1);
\node[normalnode] (x2) at (7.00, 5) {};
\draw[normalline] (node20)--(x2);
\node[normalnode] (x3) at (7.00, 4) {};
\draw[normalline] (node22)--(x3);
\node[normalnode] (x4) at (7.00, 3) {};
\draw[normalline] (node17)--(x4);
\node[normalnode] (x5) at (7.00, 2) {};
\draw[normalline] (node19)--(x5);
\node[normalnode] (x6) at (7.00, 1) {};
\draw[normalline] (node21)--(x6);
\node[normalnode] (x7) at (7.00, 0) {};
\draw[normalline] (node23)--(x7);

\end{tikzpicture}
} 
\caption{\footnotesize Relaxed factor graph} 
\label{fig:rfg}
\end{subfigure}
\caption{\footnotesize Factor graph representations of a (8,5) polar code.} \label{fig:factorgraph}
\end{figure}
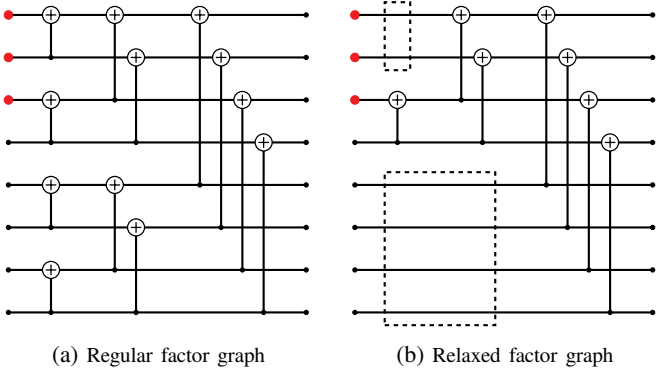

\subsection{Polar Codes}
Polar codes are a class of binary block codes with blocklength $N=2^n$ constructed from the $ n $-fold application of the polar transformation \cite{ArikanMain}, resulting in the polar transformation matrix $
    \mathbf{G}_N = \left[\begin{smallmatrix}
        1 & 0 \\
        1 & 1
    \end{smallmatrix}\right]^{\otimes n}$.
Depending on the desired rate $R$, $k$ of the $N$ inputs (so-called \textit{synthetic channels}) to the polar transformation carry the information, while the remaining $ N-k $ inputs are \textit{frozen bits} and are set to zero. The set of information bit indices is denoted by $\mathcal{A}$, while the set of frozen indices is $\mathcal{A}^c$. The selection of $\mathcal{A}$ is subject to the code design and out of scope of this paper. 

A graphical representation of polar codes is its factor graph, as depicted in Fig. \ref{fig:fg} for the (8,5) polar code with information set $\mathcal{A}=\{3,4,5,6,7\}$. Frozen bits are indicated by red dots. 

\subsection{Relaxation}
Polar code relaxation refers to the removal of vertical connections from the factor graph that do not contribute to polarization \cite{relaxedpolar}. In particular, a connection is removed if it connects either two frozen bits or two information bits. Fig.~\ref{fig:rfg} shows the relaxed factor graph of a (8,5) polar code. The removed connections are indicated by the dashed boxes.

In the case of stand-alone polar codes, relaxation does not change the set of codewords (but potentially the encoding). This means, relaxation can also be applied on only the receiver side, for example to reduce the decoding complexity or latency. Moreover, lower bit error-rates are reported in \cite{relaxedpolar}. Mathematically, we denote the relaxed polar transformation matrix as $\mathbf{G}_N^{(R)}$, which is dependent on the set of frozen bits.

\subsection{Polar Subcodes} 
While polar codes are proven to achieve the channel capacity as the block length goes to infinity ($N \to \infty$), they suffer from poor distance properties in the finite length regime. 
A common method of enhancing the performance of polar codes is to concatenate with an outer code, which removes the low-weight codewords from the polar code \cite{talvardyList}. The resulting code is a subcode of the polar code \cite{polarsubcodes}. 
A polar subcode is defined by the dynamic freezing constraint matrix $\mathbf{V} \in \mathbb{F}_2^{(N-k)\times N}$. Rows in $\mathbf{V}$ with a single `1' correspond to (hard) frozen bits, and rows with multiple `1s' are so-called dynamic frozen bits. The parity-check matrix of the concatenated code is given by
\begin{equation}
    \mathbf{H} = \mathbf{V}\cdot \mathbf{G}_N^T.
\end{equation}
Conversely, there exists a pre-transform matrix $\mathbf{W} \in \mathbb{F}_2^{k\times N}$, such that the whole polar subcode is generated by
\begin{equation}
    \mathbf{G} = \mathbf{W}\cdot \mathbf{G}_N.
\end{equation}
Note that since $\mathbf{G}_N$ is invertible, any binary linear block code of length $N$ can be viewed as a polar subcode \cite{polarsubcodes}. In the following we denote the uncoded message by $\mathbf{u}$, the pre-transformed vector by $\mathbf{v}=\mathbf{u}\mathbf{W}$ and the codeword by $\mathbf{c}=\mathbf{v}\mathbf{G}_N$.

\subsection{Affine Permutations}
In permutation-based polar decoding, codeword bits $c_i$ (and their channel \acp{LLR} $L_{\mathrm{ch},i}$) are re-arranged according to affine permutations. An affine permutation is a permutation on $ \{ 0, \dots, N-1\} $, mapping the (least significant bit first) binary representations $ \mathbf{z} $ according to
\begin{equation}
	\mathbf{z}' = \mathbf{A}\mathbf{z} + \mathbf{b} \mod 2,
\end{equation}
with $\mathbf{A}\in \mathbb{F}_2^{n\times n}$ and $\mathbf{b}\in \mathbb{F}_2^{n}$. There are three important groups of affine permutations in the context of polar codes:
\begin{enumerate}
    \item If $\mathbf{A}$ is only required to be invertible, we have the \ac{GA} group of order $n$, denoted by $\operatorname{GA}(n)$. It is well known that $\operatorname{GA}(n)$ is the automorphism group of \ac{RM} codes, i.e., the code remains the same when symbols are permuted according to any of these permutations \cite{macwilliams77}.
    \item If $\mathbf{A}$ is invertible and lower-triangular, we have the \ac{LTA} group of order $n$, denoted by $\operatorname{LTA}(n)$. $\operatorname{LTA}(n)$ is a subset of the automorphism group of polar codes \cite{bardet_polar_automorphism}.
    \item If $\mathbf{A}$ is a permutation matrix and $\mathbf{b}=\mathbf{0}$, the resulting permutation group corresponds to stage-shuffling of the polar factor graph \cite{Doan_2018_Permuted_BP}. We denote this group by $\Pi(n)$.
\end{enumerate}

\subsection{Belief Propagation Decoding of Polar Codes}
Iterative \ac{BP} decoding of polar codes is conventionally performed over the polar factor graph. In each iteration, $\mathbf{R}$-messages propagate from the left side to the right side and $\mathbf{L}$-messages vice versa through the stages of the graph. In each stage, a vertical connection in the (decoding) factor graph is called a processing element and its \ac{SISO} \ac{MAP} rule is applied in update. The $\mathbf{L}$-messages are initialized with the channel \acp{LLR} and the frozen positions of the $\mathbf{R}$-messages with saturated \acp{LLR}. Bit estimates for both $\hat{\mathbf{v}}$ and $\hat{\mathbf{c}}$ an be calculated by summing the corresponding $\mathbf{L}$- and $\mathbf{R}$-messages. A stopping condition based on 
$\mathbf{\hat{v}}\mathbf{G}_N=\mathbf{\hat{c}}$ is employed to terminate the algorithm before reaching $N_{\text{it,max}}$ iterations. Due to the \ac{SISO} nature of the algorithm, it can be connected to the Tanner graph of an outer code ``turbo-like decoder'' (see Fig.~\ref{fig:cabp}). For details on the \ac{BP} algorithm, we refer the reader to \cite{ISWCS_Error_Floor}, \cite{CRC_BPL_ISIT20}.

\subsection{Belief Propagation List Decoding}
To enhance the performance of \ac{BP} decoding, an ensemble decoder, coined \ac{BPL}, has been proposed in \cite{elkelesh2018belief}. The idea is to perform \ac{BP} on an ensemble of $L$ different stage-permuted factor graphs in parallel. Each sub-decoder contributes a codeword candidate to a list, from which the most likely valid candidate is picked as the decoder output. An extension towards CRC-aided polar codes is proposed in \cite{CRC_BPL_ISIT20}.

\section{Decoding Graphs of Polar Subcodes}
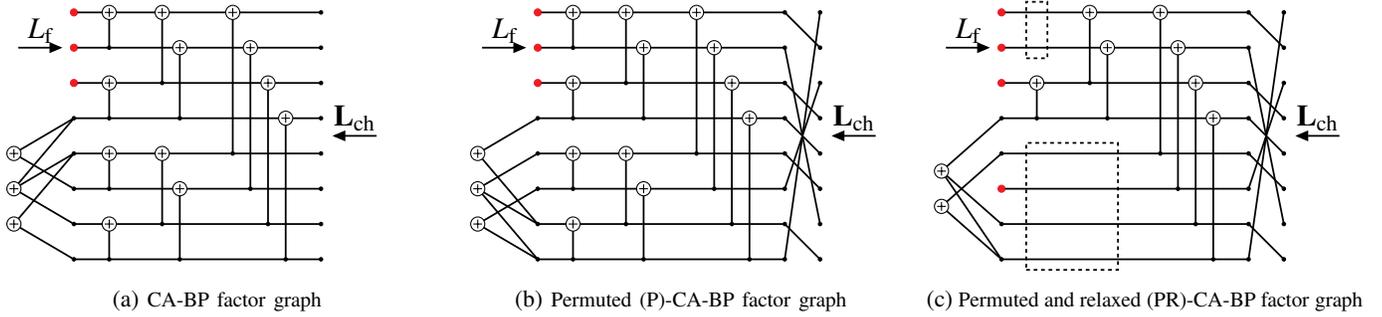
\begin{figure*}[t]

\begin{subfigure}[b]{0.32\linewidth}
\resizebox{1\columnwidth}{!}{
\begin{tikzpicture}[y=1cm]
\tikzstyle{every node}=[font=\Huge]
\tikzstyle{frozennode} = [dspnodefull,minimum size=2mm,rot]
\tikzstyle{normalnode} = [dspnodefull,minimum size=1mm]
\tikzstyle{normalline} = [line width = 0.5mm]
\tikzstyle{arrow} = [dspconn,normalline,-triangle 45]

\pgfmathtruncatemacro {\n}{3}
\pgfmathtruncatemacro {\N}{2^\n}
\pgfmathtruncatemacro {\Nm}{2^\n-1}
\pgfmathtruncatemacro {\nm}{\n-1}

\pgfmathsetmacro{\lchx}{2^\nm+1.5*\nm}
\fill [white] (-2,-.35) rectangle (10,7.35);

\node[frozennode] (u0) at (0.00, 7) {};
\node[frozennode] (u1) at (0.00, 6) {};
\node[frozennode] (u2) at (0.00, 5) {};
\node[normalnode] (u3) at (0.00, 4) {};
\node[normalnode] (u4) at (0.00, 3) {};
\node[normalnode] (u5) at (0.00, 2) {};
\node[normalnode] (u6) at (0.00, 1) {};
\node[normalnode] (u7) at (0.00, 0) {};
\node[dspadder] (node0) at (1.00, 7) {};
\node[normalnode] (node1) at (1.00, 6) {};
\draw[normalline] (u0)--(node0);
\draw[normalline] (u1)--(node1);
\draw[normalline] (node1)--(node0);
\node[dspadder] (node2) at (1.00, 5) {};
\node[normalnode] (node3) at (1.00, 4) {};
\draw[normalline] (u2)--(node2);
\draw[normalline] (u3)--(node3);
\draw[normalline] (node3)--(node2);
\node[dspadder] (node4) at (1.00, 3) {};
\node[normalnode] (node5) at (1.00, 2) {};
\draw[normalline] (u4)--(node4);
\draw[normalline] (u5)--(node5);
\draw[normalline] (node5)--(node4);
\node[dspadder] (node6) at (1.00, 1) {};
\node[normalnode] (node7) at (1.00, 0) {};
\draw[normalline] (u6)--(node6);
\draw[normalline] (u7)--(node7);
\draw[normalline] (node7)--(node6);
\node[dspadder] (node8) at (2.50, 7) {};
\node[normalnode] (node9) at (2.50, 5) {};
\draw[normalline] (node0)--(node8);
\draw[normalline] (node2)--(node9);
\draw[normalline] (node9)--(node8);
\node[dspadder] (node10) at (3.00, 6) {};
\node[normalnode] (node11) at (3.00, 4) {};
\draw[normalline] (node1)--(node10);
\draw[normalline] (node3)--(node11);
\draw[normalline] (node11)--(node10);
\node[dspadder] (node12) at (2.50, 3) {};
\node[normalnode] (node13) at (2.50, 1) {};
\draw[normalline] (node4)--(node12);
\draw[normalline] (node6)--(node13);
\draw[normalline] (node13)--(node12);
\node[dspadder] (node14) at (3.00, 2) {};
\node[normalnode] (node15) at (3.00, 0) {};
\draw[normalline] (node5)--(node14);
\draw[normalline] (node7)--(node15);
\draw[normalline] (node15)--(node14);
\node[dspadder] (node16) at (4.50, 7) {};
\node[normalnode] (node17) at (4.50, 3) {};
\draw[normalline] (node8)--(node16);
\draw[normalline] (node12)--(node17);
\draw[normalline] (node17)--(node16);
\node[dspadder] (node18) at (5.00, 6) {};
\node[normalnode] (node19) at (5.00, 2) {};
\draw[normalline] (node10)--(node18);
\draw[normalline] (node14)--(node19);
\draw[normalline] (node19)--(node18);
\node[dspadder] (node20) at (5.50, 5) {};
\node[normalnode] (node21) at (5.50, 1) {};
\draw[normalline] (node9)--(node20);
\draw[normalline] (node13)--(node21);
\draw[normalline] (node21)--(node20);
\node[dspadder] (node22) at (6.00, 4) {};
\node[normalnode] (node23) at (6.00, 0) {};
\draw[normalline] (node11)--(node22);
\draw[normalline] (node15)--(node23);
\draw[normalline] (node23)--(node22);
\node[normalnode] (x0) at (7.00, 7) {};
\draw[normalline] (node16)--(x0);
\node[normalnode] (x1) at (7.00, 6) {};
\draw[normalline] (node18)--(x1);
\node[normalnode] (x2) at (7.00, 5) {};
\draw[normalline] (node20)--(x2);
\node[normalnode] (x3) at (7.00, 4) {};
\draw[normalline] (node22)--(x3);
\node[normalnode] (x4) at (7.00, 3) {};
\draw[normalline] (node17)--(x4);
\node[normalnode] (x5) at (7.00, 2) {};
\draw[normalline] (node19)--(x5);
\node[normalnode] (x6) at (7.00, 1) {};
\draw[normalline] (node21)--(x6);
\node[normalnode] (x7) at (7.00, 0) {};
\draw[normalline] (node23)--(x7);

\node[] (ch2) at (\lchx+.2, \Nm/2) {};
\node[] (ch1) at (\lchx+1.7, \Nm/2) {};
\draw[arrow](ch1)--node[above] {$\Lm_{\mathrm{ch}}$}(ch2);
	
\node[] (src1) at (-1.7, 6) {};
\node[] (src2) at (-0.2, 6) {};
\draw[arrow](src1)--node[above] {$L_{\mathrm{f}}$}(src2);



\node[dspadder]    (xorcrc1) at (-1.7, 3) {};
\node[dspadder]    (xorcrc2) at (-1.7, 2) {};
\node[dspadder]    (xorcrc3) at (-1.7, 1) {};

\draw[normalline] (u3)--(xorcrc1);
\draw[normalline] (u5)--(xorcrc1);

\draw[normalline] (u3)--(xorcrc2);
\draw[normalline] (u4)--(xorcrc2);
\draw[normalline] (u6)--(xorcrc2);

\draw[normalline] (u4)--(xorcrc3);
\draw[normalline] (u7)--(xorcrc3);


\end{tikzpicture}
} 
\caption{\footnotesize CA-BP factor graph} 
\label{fig:cabp}
\end{subfigure}
\hfill     
\begin{subfigure}[b]{0.32\linewidth}
\resizebox{1\columnwidth}{!}{
\begin{tikzpicture}[y=1cm]
\tikzstyle{every node}=[font=\Huge]
\tikzstyle{frozennode} = [dspnodefull,minimum size=2mm,rot]
\tikzstyle{normalnode} = [dspnodefull,minimum size=1mm]
\tikzstyle{normalline} = [line width = 0.5mm]
\tikzstyle{arrow} = [dspconn,normalline,-triangle 45]

\pgfmathtruncatemacro {\n}{3}
\pgfmathtruncatemacro {\N}{2^\n}
\pgfmathtruncatemacro {\Nm}{2^\n-1}
\pgfmathtruncatemacro {\nm}{\n-1}

\pgfmathsetmacro{\lchx}{2^\nm+1.5*\nm+1}
\fill [white] (-2,-.35) rectangle (10,7.35);

\foreach \b in {0,1,2,3,4,5,6,7}
{
	\node[normalnode] (lch\b) at (\lchx, \b) {};
}

\node[frozennode] (u0) at (0.00, 7) {};
\node[frozennode] (u1) at (0.00, 6) {};
\node[frozennode] (u2) at (0.00, 5) {};
\node[normalnode] (u3) at (0.00, 4) {};
\node[normalnode] (u4) at (0.00, 3) {};
\node[normalnode] (u5) at (0.00, 2) {};
\node[normalnode] (u6) at (0.00, 1) {};
\node[normalnode] (u7) at (0.00, 0) {};
\node[dspadder] (node0) at (1.00, 7) {};
\node[normalnode] (node1) at (1.00, 6) {};
\draw[normalline] (u0)--(node0);
\draw[normalline] (u1)--(node1);
\draw[normalline] (node1)--(node0);
\node[dspadder] (node2) at (1.00, 5) {};
\node[normalnode] (node3) at (1.00, 4) {};
\draw[normalline] (u2)--(node2);
\draw[normalline] (u3)--(node3);
\draw[normalline] (node3)--(node2);
\node[dspadder] (node4) at (1.00, 3) {};
\node[normalnode] (node5) at (1.00, 2) {};
\draw[normalline] (u4)--(node4);
\draw[normalline] (u5)--(node5);
\draw[normalline] (node5)--(node4);
\node[dspadder] (node6) at (1.00, 1) {};
\node[normalnode] (node7) at (1.00, 0) {};
\draw[normalline] (u6)--(node6);
\draw[normalline] (u7)--(node7);
\draw[normalline] (node7)--(node6);
\node[dspadder] (node8) at (2.50, 7) {};
\node[normalnode] (node9) at (2.50, 5) {};
\draw[normalline] (node0)--(node8);
\draw[normalline] (node2)--(node9);
\draw[normalline] (node9)--(node8);
\node[dspadder] (node10) at (3.00, 6) {};
\node[normalnode] (node11) at (3.00, 4) {};
\draw[normalline] (node1)--(node10);
\draw[normalline] (node3)--(node11);
\draw[normalline] (node11)--(node10);
\node[dspadder] (node12) at (2.50, 3) {};
\node[normalnode] (node13) at (2.50, 1) {};
\draw[normalline] (node4)--(node12);
\draw[normalline] (node6)--(node13);
\draw[normalline] (node13)--(node12);
\node[dspadder] (node14) at (3.00, 2) {};
\node[normalnode] (node15) at (3.00, 0) {};
\draw[normalline] (node5)--(node14);
\draw[normalline] (node7)--(node15);
\draw[normalline] (node15)--(node14);
\node[dspadder] (node16) at (4.50, 7) {};
\node[normalnode] (node17) at (4.50, 3) {};
\draw[normalline] (node8)--(node16);
\draw[normalline] (node12)--(node17);
\draw[normalline] (node17)--(node16);
\node[dspadder] (node18) at (5.00, 6) {};
\node[normalnode] (node19) at (5.00, 2) {};
\draw[normalline] (node10)--(node18);
\draw[normalline] (node14)--(node19);
\draw[normalline] (node19)--(node18);
\node[dspadder] (node20) at (5.50, 5) {};
\node[normalnode] (node21) at (5.50, 1) {};
\draw[normalline] (node9)--(node20);
\draw[normalline] (node13)--(node21);
\draw[normalline] (node21)--(node20);
\node[dspadder] (node22) at (6.00, 4) {};
\node[normalnode] (node23) at (6.00, 0) {};
\draw[normalline] (node11)--(node22);
\draw[normalline] (node15)--(node23);
\draw[normalline] (node23)--(node22);
\node[normalnode] (x0) at (7.00, 7) {};
\draw[normalline] (node16)--(x0);
\node[normalnode] (x1) at (7.00, 6) {};
\draw[normalline] (node18)--(x1);
\node[normalnode] (x2) at (7.00, 5) {};
\draw[normalline] (node20)--(x2);
\node[normalnode] (x3) at (7.00, 4) {};
\draw[normalline] (node22)--(x3);
\node[normalnode] (x4) at (7.00, 3) {};
\draw[normalline] (node17)--(x4);
\node[normalnode] (x5) at (7.00, 2) {};
\draw[normalline] (node19)--(x5);
\node[normalnode] (x6) at (7.00, 1) {};
\draw[normalline] (node21)--(x6);
\node[normalnode] (x7) at (7.00, 0) {};
\draw[normalline] (node23)--(x7);
\def\permutation{{1,6,3,4,5,2,7,0}}
\foreach \i in {0,1,2,3,4,5,6,7}
	\draw[normalline] (\lchx-1,\permutation[\i]) -- (lch\i);

\node[] (ch2) at (\lchx+.2, \Nm/2) {};
\node[] (ch1) at (\lchx+1.7, \Nm/2) {};
\draw[arrow](ch1)--node[above] {$\Lm_{\mathrm{ch}}$}(ch2);
	
\node[] (src1) at (-1.7, 6) {};
\node[] (src2) at (-0.2, 6) {};
\draw[arrow](src1)--node[above] {$L_{\mathrm{f}}$}(src2);



\node[dspadder]    (xorcrc1) at (-1.7, 3) {};
\node[dspadder]    (xorcrc2) at (-1.7, 2) {};
\node[dspadder]    (xorcrc3) at (-1.7, 1) {};

\draw[normalline] (u3)--(xorcrc1);
\draw[normalline] (u6)--(xorcrc1);

\draw[normalline] (u4)--(xorcrc2);
\draw[normalline] (u6)--(xorcrc2);
\draw[normalline] (u7)--(xorcrc2);

\draw[normalline] (u5)--(xorcrc3);
\draw[normalline] (u7)--(xorcrc3);


\end{tikzpicture}
} 
\caption{\footnotesize Permuted (P)-CA-BP factor graph} 
\label{fig:pcabp}
\end{subfigure}
\hfill     
\begin{subfigure}[b]{0.32\linewidth}
\resizebox{1\columnwidth}{!}{
\begin{tikzpicture}[y=1cm]
\tikzstyle{every node}=[font=\Huge]
\tikzstyle{frozennode} = [dspnodefull,minimum size=2mm,rot]
\tikzstyle{normalnode} = [dspnodefull,minimum size=1mm]
\tikzstyle{normalline} = [line width = 0.5mm]
\tikzstyle{arrow} = [dspconn,normalline,-triangle 45]

\pgfmathtruncatemacro {\n}{3}
\pgfmathtruncatemacro {\N}{2^\n}
\pgfmathtruncatemacro {\Nm}{2^\n-1}
\pgfmathtruncatemacro {\nm}{\n-1}

\pgfmathsetmacro{\lchx}{2^\nm+1.5*\nm+1}

\fill [white] (-2,-.35) rectangle (10,7.35);

\foreach \b in {0,1,2,3,4,5,6,7}
{
	\node[normalnode] (lch\b) at (\lchx, \b) {};
}

\node[frozennode] (u0) at (0.00, 7) {};
\node[frozennode] (u1) at (0.00, 6) {};
\node[frozennode] (u2) at (0.00, 5) {};
\node[normalnode] (u3) at (0.00, 4) {};
\node[normalnode] (u4) at (0.00, 3) {};
\node[frozennode] (u5) at (0.00, 2) {};
\node[normalnode] (u6) at (0.00, 1) {};
\node[normalnode] (u7) at (0.00, 0) {};

\draw[style=dashed, line width = .5mm, black] (0.7,5.7) rectangle (1.3,7.3);
\node[dspadder] (node2) at (1.00, 5) {};
\node[normalnode] (node3) at (1.00, 4) {};
\draw[normalline] (u2)--(node2);
\draw[normalline] (u3)--(node3);
\draw[normalline] (node3)--(node2);
\draw[style=dashed, line width = .5mm, black] (0.7,-0.3) rectangle (3.3,3.3);
\node[dspadder] (node8) at (2.50, 7) {};
\node[normalnode] (node9) at (2.50, 5) {};
\draw[normalline] (u0)--(node8);
\draw[normalline] (node2)--(node9);
\draw[normalline] (node9)--(node8);
\node[dspadder] (node10) at (3.00, 6) {};
\node[normalnode] (node11) at (3.00, 4) {};
\draw[normalline] (u1)--(node10);
\draw[normalline] (node3)--(node11);
\draw[normalline] (node11)--(node10);
\node[dspadder] (node16) at (4.50, 7) {};
\node[normalnode] (node17) at (4.50, 3) {};
\draw[normalline] (node8)--(node16);
\draw[normalline] (u4)--(node17);
\draw[normalline] (node17)--(node16);
\node[dspadder] (node18) at (5.00, 6) {};
\node[normalnode] (node19) at (5.00, 2) {};
\draw[normalline] (node10)--(node18);
\draw[normalline] (u5)--(node19);
\draw[normalline] (node19)--(node18);
\node[dspadder] (node20) at (5.50, 5) {};
\node[normalnode] (node21) at (5.50, 1) {};
\draw[normalline] (node9)--(node20);
\draw[normalline] (u6)--(node21);
\draw[normalline] (node21)--(node20);
\node[dspadder] (node22) at (6.00, 4) {};
\node[normalnode] (node23) at (6.00, 0) {};
\draw[normalline] (node11)--(node22);
\draw[normalline] (u7)--(node23);
\draw[normalline] (node23)--(node22);
\node[normalnode] (x0) at (7.00, 7) {};
\draw[normalline] (node16)--(x0);
\node[normalnode] (x1) at (7.00, 6) {};
\draw[normalline] (node18)--(x1);
\node[normalnode] (x2) at (7.00, 5) {};
\draw[normalline] (node20)--(x2);
\node[normalnode] (x3) at (7.00, 4) {};
\draw[normalline] (node22)--(x3);
\node[normalnode] (x4) at (7.00, 3) {};
\draw[normalline] (node17)--(x4);
\node[normalnode] (x5) at (7.00, 2) {};
\draw[normalline] (node19)--(x5);
\node[normalnode] (x6) at (7.00, 1) {};
\draw[normalline] (node21)--(x6);
\node[normalnode] (x7) at (7.00, 0) {};
\draw[normalline] (node23)--(x7);

\def\permutation{{1,6,3,4,5,2,7,0}}
\foreach \i in {0,1,2,3,4,5,6,7}
	\draw[normalline] (\lchx-1,\permutation[\i]) -- (lch\i);

\node[] (ch2) at (\lchx+.2, \Nm/2) {};
\node[] (ch1) at (\lchx+1.7, \Nm/2) {};
\draw[arrow](ch1)--node[above] {$\Lm_{\mathrm{ch}}$}(ch2);
	
\node[] (src1) at (-1.7, 6) {};
\node[] (src2) at (-0.2, 6) {};
\draw[arrow](src1)--node[above] {$L_{\mathrm{f}}$}(src2);



\node[dspadder]    (xorcrc1) at (-1.7, 2.5) {};
\node[dspadder]    (xorcrc2) at (-1.7, 1.5) {};

\draw[normalline] (u3)--(xorcrc1);
\draw[normalline] (u6)--(xorcrc1);
\draw[normalline] (u7)--(xorcrc1);

\draw[normalline] (u4)--(xorcrc2);
\draw[normalline] (u7)--(xorcrc2);


\end{tikzpicture}
} 
\caption{\footnotesize Permuted and relaxed (PR)-CA-BP factor graph} 
\label{fig:prcabp}
\end{subfigure}
\caption{\footnotesize Different factor graph representations of a (8,5) polar code concatenated with an outer CRC-3 code with $g(x)=x^3+x+1$. Note that the outer code factor graph is different in all three cases.} \label{fig:graphs}
\end{figure*}
In this section we derive methods for obtaining different factor graphs of polar subcodes for ensemble decoding based on affine permutations and relaxation.

\subsection{Post-Transformations}
Given an invertible binary $N\times N$  matrix $\mathbf{T}$, we can write the generator matrix of any polar subcode as
\begin{equation}
\mathbf{W} \mathbf{G}_N \mathbf{T} \mathbf{T}^{-1} = \mathbf{W} \mathbf{G}_N \mathbf{T} \mathbf{G}_N^{-1} \mathbf{G}_N
\mathbf{T}^{-1} = \mathbf{W}_T \mathbf{G}_N \mathbf{T}^{-1}.
\end{equation}
In other words, the encoding is equal to that of a different polar subcode followed by the post-transformation $\mathbf{T}^{-1}$. The pre-transform matrix and dynamic freezing constraint matrix of the new polar subcode are 
\begin{align}
\mathbf{W}_T &= \mathbf{W} \mathbf{G}_N \mathbf{T}\mathbf{G}_N \quad \text{and}\\
\mathbf{V}_T &= \mathbf{V} \left(\mathbf{G}_N \mathbf{T}^{-1}\mathbf{G}_N\right)^\mathrm{T},
\end{align}
respectively, where we used $\mathbf{G}_N^{-1}=\mathbf{G}_N$. A particular interesting case is when $\mathbf{T}$ is a permutation matrix, as this can be easily implemented at the first decoding stage as a permutation of the received sequence.

\textbf{Example 1:}
Fig. \ref{fig:cabp} shows the factor graph of a (8,5) polar code concatenated with a 3-bit \ac{CRC} (i.e., CRC-3) described by the generator polynomial $g(x)=x^3+x+1$. It can be interpreted as a polar subcode with dynamic freezing constraint matrix
\begin{equation}
    \mathbf{V} = \left[\begin{array}{cccccccc}
1&0&0&0&0&0&0&0\\
0&1&0&0&0&0&0&0\\
0&0&1&0&0&0&0&0\\
0&0&0&1&0&1&0&0\\
0&0&0&1&1&0&1&0\\
0&0&0&0&1&0&0&1
\end{array}\right].
\end{equation}
Observe that the first three rows correspond to the frozen bits, while the remaining three rows contain the parity-check matrix of the \ac{CRC} code. 
Now, consider the \ac{LTA} permutation $\pi=(1,6,3,4,5,2,7,0)$ corresponding to the pair 
\begin{equation}
    \mathbf{A} = \left[\begin{array}{ccc}
1&0&0\\
1&1&0\\
1&0&1
\end{array}\right], \; \mathbf{b} = \left[\begin{array}{c}
1\\
0\\
0
\end{array}\right].
\end{equation}
Thus, its post-transformation matrix is
\begin{equation}
    \mathbf{T}^{-1} = \left[\begin{array}{cccccccc}
0&0&0&0&0&0&0&1\\
1&0&0&0&0&0&0&0\\
0&0&0&0&0&1&0&0\\
0&0&1&0&0&0&0&0\\
0&0&0&1&0&0&0&0\\
0&0&0&0&1&0&0&0\\
0&1&0&0&0&0&0&0\\
0&0&0&0&0&0&1&0
\end{array}\right].
\end{equation}
We then find the transformed dynamic freezing constraint matrix
\begin{equation}
    \mathbf{V}_T = \mathbf{V} \left(\mathbf{G}_N \mathbf{T}^{-1}\mathbf{G}_N\right)^\mathrm{T} = \left[\begin{array}{cccccccc}
1&0&0&0&0&0&0&0\\
1&1&0&0&0&0&0&0\\
0&1&1&0&0&0&0&0\\
0&0&1&1&1&1&0&0\\
0&0&1&0&1&1&1&0\\
0&1&0&0&1&0&1&1
\end{array}\right].\label{eq:vt}
\end{equation}
Note that the first three columns remain frozen, which can be easily seen after a few row operations. Moreover, additional row-operations can be applied to the dynamic frozen bits, reducing the number of edges in the factor graph, as depicted in Fig. \ref{fig:pcabp}. The ensemble variant of this decoder is denoted by \ac{P-CA-BPL}-$L$ decoding.

\subsection{Polar Subcode Relaxation}
If the same outer code is used together with a relaxed polar code, the resulting concatenated code is generally different than without relaxation. Hence, to comply with a non-relaxed encoder, the outer code must be altered to account for relaxation. The idea is to include the relaxed (i.e., removed) partial polar transformations in the outer code. Those removed polar transformations can be collected in the matrix $\mathbf{R}$, fulfilling
\begin{equation}
 \mathbf{R} \cdot \mathbf{G}_N^{(R)} = \mathbf{G}_N.
\end{equation}
For a given dynamic freezing constraint matrix $\mathbf{V}$, the equivalent matrix for the relaxed polar code equates to
\begin{equation}
 \mathbf{V}_R = \mathbf{V} \cdot \mathbf{R}^{\mathrm{T}}. \label{eq:relax}
\end{equation}
Similarly, $\mathbf{R}$ can also be applied to a permuted polar code, i.e., its  matrix $ \mathbf{V}_T$. 
It should be noted that this variation of relaxation in general (i.e., for outer codes with dense parity-check matrix) retain the decoding complexity reduction as experienced in stand-alone polar codes, as the outer code usually remains high-density when multiplying by $\mathbf{R}$.

\textbf{Example 1 (cont'd):}
As the inner polar code is the same as in Fig. \ref{fig:factorgraph}, relaxation removes the same processing elements. We find the relaxation matrix 
\begin{equation}
    \mathbf{R}= \left[\begin{array}{cccccccc}
1&0&0&0&0&0&0&0\\
1&1&0&0&0&0&0&0\\
0&0&1&0&0&0&0&0\\
0&0&0&1&0&0&0&0\\
0&0&0&0&1&0&0&0\\
0&0&0&0&1&1&0&0\\
0&0&0&0&1&0&1&0\\
0&0&0&0&1&1&1&1
\end{array}\right].
\end{equation}
Using Eq.~(\ref{eq:relax}) and a few row operations, we find the permuted relaxed dynamic freezing constraint matrix for (\ref{eq:vt}) to be
\begin{equation}
    \mathbf{V}_{T,R} = \left[\begin{array}{cccccccc}
1&0&0&0&0&0&0&0\\
0&1&0&0&0&0&0&0\\
0&0&1&0&0&0&0&0\\
0&0&0&0&0&1&0&0\\
0&0&0&1&0&0&1&1\\
0&0&0&0&1&0&0&1
\end{array}\right].\label{eq:vtr}
\end{equation}
Note that bit 5 turned from a dynamic frozen bit into a hard frozen bit. Fig. \ref{fig:prcabp} shows the permuted relaxed factor graph. 

A decoder using this factor graph is called \ac{R-CA-BP}, and its ensemble variant \ac{R-CA-BPL}-$L$. If combined with an affine permutation, we refer to it as \ac{PR-CA-BPL}-$L$ decoding.

\section{Results}
For all shown results, we selected the information/frozen-bit positions according to the 5G standard \cite{polar5G2018} and the outer code is a CRC-8 described by the generator polynomial $g(x) = x^8 + x^5 + x^4 + x^3 + 1$. Similar to \cite{CRC_BPL_ISIT20}, we apply a heuristic density reduction to the dynamic freezing constraint matrices to be better suited for the \ac{SPA}. For all shown ensemble decoders, we do not optimize the set of permutations that are used, i.e., we select random permutations from the respective groups. However, optimization methods (e.g., genetic algorithm-based as proposed in \cite{CRC_BPL_ISIT20}) are applicable and expected to provide additional gains. In all simulations, we assume that the codeword is \ac{BPSK} modulated and transmitted over an \ac{AWGN} channel.

\subsection{Permutation Subgroups}
\begin{figure} [t]
	\centering
	\resizebox{\columnwidth}{!}{\begin{tikzpicture}
\begin{axis}[
width=\linewidth,
height=.58\linewidth,
grid style={dotted,anthrazit},
xmajorgrids,
yminorticks=true,
ymajorgrids,
legend columns=1,
legend pos=south west,   
legend cell align={left},
legend style={fill,fill opacity=0.8},
xlabel={$E_\mathrm{b}/N_0$ [dB]},
ylabel={BLER},
legend image post style={mark indices={}},
ymode=log,
mark size=1.5pt,
xmin=1,
xmax=4,
ymin=1.3e-03,
ymax=7.299e-01,
]

\addplot[color=magenta,line width = 1pt, solid,mark=square,mark size=2.5pt, mark options={solid}]
table[col sep=comma]{
0.00, 9.174e-01
0.50, 9.009e-01
1.00, 7.299e-01
1.50, 5.319e-01
2.00, 3.460e-01
2.50, 1.996e-01
3.00, 8.220e-02
3.50, 2.990e-02
4.00, 7.945e-03
};
\label{plot:ga}
\addlegendentry{\footnotesize GA permutations};

\addplot[color=apfelgruen,line width = 1pt, solid,mark=x,mark size=2.5pt, mark options={solid}]
table[col sep=comma]{
0.00, 9.560e-01
0.50, 8.881e-01
1.00, 7.215e-01
1.50, 4.721e-01
2.00, 2.726e-01
2.50, 1.279e-01
3.00, 4.998e-02
3.50, 1.399e-02
4.00, 3.551e-03
};
\label{plot:pi}
\addlegendentry{\footnotesize $\Pi$ permutations};

\addplot[color=mittelblau,line width = 1pt, solid,mark=triangle,mark size=2.5pt, mark options={solid}]
table[col sep=comma]{
0.00, 9.363e-01
0.50, 8.306e-01
1.00, 6.859e-01
1.50, 4.163e-01
2.00, 2.288e-01
2.50, 8.794e-02
3.00, 2.724e-02
3.50, 7.070e-03
4.00, 1.572e-03
};
\label{plot:lta}
\addlegendentry{\footnotesize LTA permutations};

\addplot[color=gray,line width = 1pt, solid,mark size=2.5pt, mark options={solid}]
table[col sep=comma]{
1.00, 6.465e-01
1.50, 4.399e-01
2.00, 2.331e-01
2.50, 1.026e-01
3.00, 3.075e-02
3.50, 6.973e-03
4.00, 1.630e-03
};
\label{plot:def}
\addlegendentry{\footnotesize default factor graph};

\end{axis}

\end{tikzpicture}}
	\caption{\footnotesize Error-rate performance of a single \ac{PR-CA-BP} decoder using 100 iterations for the (128,64)-Polar-CRC8 code, random permutations.}
	\label{fig:permutations_results}
\end{figure}
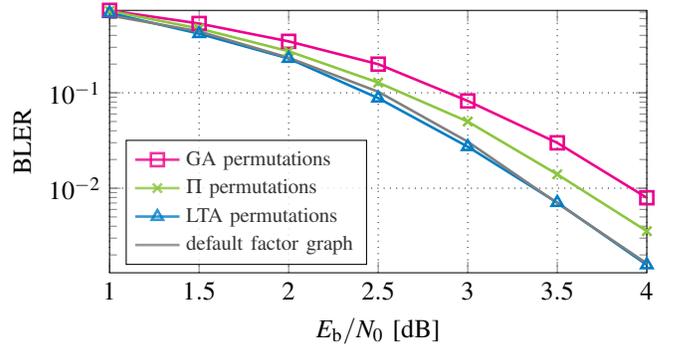

While in \cite{KamenevPermutedSCwithAutomorphisms} it is shown that permutations from all subsets of the \ac{GA} group does not degrade the distance of the transformed code, the performance of the iterative decoding scheme is still highly dependent on the type of permutation. This is due to the changing number and position of the hard frozen bits (i.e., the effective polar code design) and the transformed dynamic frozen bits. To illustrate this, we plot the average performance of a single \acf{PR-CA-BP} decoder using different permutations in Fig. \ref{fig:permutations_results}. For every simulated frame, a new random permutation is selected from each permutation subgroup. As we can see, the \ac{LTA} permuted decoder has the best performance, similar to the default (i.e., no) permutation. Stage shuffle permutations $\Pi$ deteriorate the performance on average by 0.2 dB at a \ac{BLER} of $10^{-2}$, and \ac{GA} permutations lose on average another 0.3 dB. This effect can be explained by the fact that \ac{LTA} permutations, which are naturally automorphisms of the inner polar code, do not change the set of frozen/information bits. Hence, we expect all \ac{LTA} permuted decoders to perform similarly to the unpermuted decoder, while permutations which are not automorphisms may deteriorate the effective polar code design. Based on this result, it is reasonable to limit the permutations used in the ensemble/list decoder to \ac{LTA} permutations.\footnote{The limitation to \ac{LTA} is for the sake of simplicity; the same argument holds for the full automorphism group of polar codes \cite{PolarAutomorphisms_ISIT21,LiBLTA2021}.}

\subsection{Ensemble Decoding Error-Rate Performance}
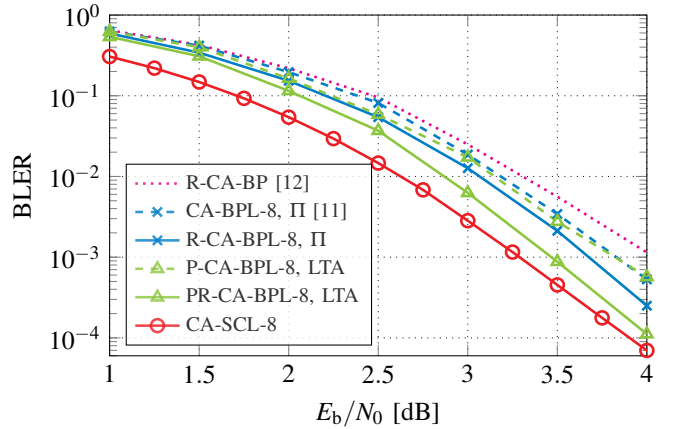
\begin{figure} [t]
	\centering
	\resizebox{\columnwidth}{!}{\begin{tikzpicture}
\begin{axis}[
width=\linewidth,
height=.7\linewidth,
grid style={dotted,anthrazit},
xmajorgrids,
yminorticks=true,
ymajorgrids,
legend columns=1,
legend pos=south west,   
legend cell align={left},
legend style={fill,fill opacity=0.8},
xlabel={$E_\mathrm{b}/N_0$ [dB]},
ylabel={BLER},
legend image post style={mark indices={}},
ymode=log,
mark size=1.5pt,
xmin=1,
xmax=4,
ymin=6e-5,
ymax=1
]

\addplot[color=magenta,line width = 1pt, dotted,mark=none,mark size=2.5pt, mark options={solid}]
table[col sep=comma]{
1.00, 6.465e-01
1.50, 4.271e-01
2.00, 2.190e-01
2.50, 9.529e-02
3.00, 2.493e-02
3.50, 5.616e-03
4.00, 1.144e-03
};
\label{plot:128cabp}
\addlegendentry{\footnotesize R-CA-BP \cite{relaxedcrc}};

\addplot[color=mittelblau,line width = 1pt, dashed,mark=x,mark size=2.5pt, mark options={solid}]
table[col sep=comma]{
1.00, 6.265e-01
1.50, 4.180e-01
2.00, 1.963e-01
2.50, 8.143e-02
3.00, 1.868e-02
3.50, 3.427e-03
4.00, 5.307e-04
};
\label{plot:128cabpl8}
\addlegendentry{\footnotesize CA-BPL-8, $\Pi$ \cite{CRC_BPL_ISIT20}};

\addplot[color=mittelblau,line width = 1pt, solid,mark=x,mark size=2.5pt, mark options={solid}]
table[col sep=comma]{
1.00, 5.912e-01
1.50, 3.419e-01
2.00, 1.534e-01
2.50, 5.424e-02
3.00, 1.267e-02
3.50, 2.120e-03
4.00, 2.503e-04
};
\label{plot:128cabpl8}
\addlegendentry{\footnotesize R-CA-BPL-8, $\Pi$};

\addplot[color=apfelgruen,line width = 1pt, dashed,mark=triangle,mark size=2.5pt, mark options={solid}]
table[col sep=comma]{
1.00, 6.243e-01
1.50, 4.000e-01
2.00, 1.636e-01
2.50, 5.856e-02
3.00, 1.728e-02
3.50, 2.759e-03
4.00, 5.691e-04
};
\label{plot:128cabpl8}
\addlegendentry{\footnotesize P-CA-BPL-8, LTA};

\addplot[color=apfelgruen,line width = 1pt, solid,mark=triangle,mark size=2.5pt, mark options={solid}]
table[col sep=comma]{
1.00, 5.381e-01
1.50, 3.086e-01
2.00, 1.149e-01
2.50, 3.689e-02
3.00, 6.267e-03
3.50, 8.786e-04
4.00, 1.115e-04
};
\label{plot:128cabpl8}
\addlegendentry{\footnotesize PR-CA-BPL-8, LTA};

\addplot[color=rot,line width = 1pt, solid,mark=o,mark size=2.5pt, mark options={solid}]
table[col sep=comma]{
1.00, 3.061e-01
1.25, 2.199e-01
1.50, 1.481e-01
1.75, 9.288e-02
2.00, 5.429e-02
2.25, 2.943e-02
2.50, 1.465e-02
2.75, 6.824e-03
3.00, 2.831e-03
3.25, 1.156e-03
3.50, 4.530e-04
3.75, 1.770e-04
4.00, 6.990e-05
4.25, 2.600e-05
4.50, 1.700e-05
4.75, 6.000e-06
};
\label{plot:128scl8}
\addlegendentry{\footnotesize CA-SCL-8};

\end{axis}

\end{tikzpicture}}
	\caption{\footnotesize Error-rate performance of BPL and SCL decoding for the (128,64)-Polar-CRC8 code. All iterative decoders use $N_\mathrm{it,max}=200$ iterations.}
	\label{fig:n128}
\end{figure}
\begin{figure} [t]
	\centering
	\resizebox{\columnwidth}{!}{\begin{tikzpicture}
\begin{axis}[
width=\linewidth,
height=.7\linewidth,
grid style={dotted,anthrazit},
xmajorgrids,
yminorticks=true,
ymajorgrids,
legend columns=1,
legend pos=south west,   
legend cell align={left},
legend style={fill,fill opacity=0.8},
xlabel={$E_\mathrm{b}/N_0$ [dB]},
ylabel={BLER},
legend image post style={mark indices={}},
ymode=log,
mark size=1.5pt,
xmin=1,
xmax=3.5,
ymin=1e-5,
ymax=1
]

\addplot[color=magenta,line width = 1pt, dotted,mark=none,mark size=2.5pt, mark options={solid}]
table[col sep=comma]{
1.00, 6.184e-01
1.50, 3.342e-01
2.00, 1.241e-01
2.50, 2.744e-02
3.00, 3.882e-03
3.50, 4.710e-04
4.00, 5.788e-05
};
\label{plot:128cabp}
\addlegendentry{\footnotesize R-CA-BP \cite{relaxedcrc}};

\addplot[color=mittelblau,line width = 1pt, dashed,mark=x,mark size=2.5pt, mark options={solid}]
table[col sep=comma]{
1.00, 6.174e-01
1.50, 3.187e-01
2.00, 1.003e-01
2.50, 2.036e-02
3.00, 2.415e-03
3.50, 2.665e-04
4.00, 2.536e-05
};
\label{plot:128cabpl8}
\addlegendentry{\footnotesize CA-BPL-8, $\Pi$ \cite{CRC_BPL_ISIT20}};

\addplot[color=mittelblau,line width = 1pt, solid,mark=x,mark size=2.5pt, mark options={solid}]
table[col sep=comma]{
1.00, 5.753e-01
1.50, 2.851e-01
2.00, 8.779e-02
2.50, 1.598e-02
3.00, 1.598e-03
3.50, 1.482e-04
};
\label{plot:128cabpl8}
\addlegendentry{\footnotesize R-CA-BPL-8, $\Pi$};

\addplot[color=apfelgruen,line width = 1pt, dashed,mark=triangle,mark size=2.5pt, mark options={solid}]
table[col sep=comma]{
1.00, 5.347e-01
1.50, 2.280e-01
2.00, 6.235e-02
2.50, 1.175e-02
3.00, 1.653e-03
3.50, 1.122e-04
4.00, 1.310e-05
};
\label{plot:256cabpl8}
\addlegendentry{\footnotesize P-CA-BPL-8, LTA};

\addplot[color=apfelgruen,line width = 1pt, solid,mark=triangle,mark size=2.5pt, mark options={solid}]
table[col sep=comma]{
1.00, 4.894e-01
1.50, 2.210e-01
2.00, 5.300e-02
2.50, 8.278e-03
3.00, 7.181e-04
3.50, 4.888e-05
};
\label{plot:128cabpl8}
\addlegendentry{\footnotesize PR-CA-BPL-8, LTA};

\addplot[color=rot,line width = 1pt, solid,mark=o,mark size=2.5pt, mark options={solid}]
table[col sep=comma]{
1.00, 2.747e-01
1.25, 1.699e-01
1.50, 9.417e-02
1.75, 4.661e-02
2.00, 2.038e-02
2.25, 7.798e-03
2.50, 2.666e-03
2.75, 7.830e-04
3.00, 2.020e-04
3.25, 4.600e-05
3.50, 1.000e-05
};
\label{plot:256scl8}
\addlegendentry{\footnotesize CA-SCL-8};

\end{axis}

\end{tikzpicture}}
	\caption{\footnotesize Error-rate performance of BPL and SCL decoding for the (256,128)-Polar-CRC8 code. All iterative decoders use $N_\mathrm{it,max}=200$ iterations.}
	\label{fig:n256}
\end{figure}
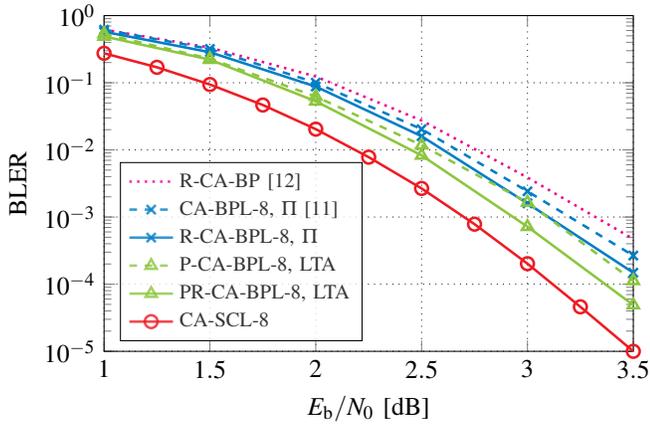
Fig. \ref{fig:n128} compares the \ac{BLER} performance of different iterative decoding algorithms with state-of-the-art \ac{SCL} decoding ($L=8$) for the (128,64) \ac{CRC}-aided polar code. All ensemble decoders use $L=8$ constituent decoders. At a \ac{BLER} of $10^{-3}$, the proposed \ac{PR-CA-BPL} scheme outperforms a single \ac{R-CA-BP} decoder by more than 0.5 dB. Compared to \ac{CA-BPL} (with stage shuffle permutations), the richer \ac{LTA} permutations result in a gain of 0.4 dB. Roughly half of this gain is attributed to the use of relaxation. Interestingly, relaxation has a greater effect on \ac{LTA} permutations than on $\Pi$ permutations, as their un-relaxed counter parts have equal performance. Lastly, \ac{PR-CA-BPL} closes the gap to \ac{SCL} decoding to 0.1 dB at a \ac{BLER} of $10^{-4}$.
In Fig. \ref{fig:n256}, we show the same comparison for the longer, (256,128) \ac{CRC}-aided polar code. 
We see similar results as for $N=128$ with a gain of ensemble decoding over \ac{R-CA-BP} decoding of roughly 0.4 dB.

\subsection{Latency}
\begin{figure} [t]
	\centering
	\resizebox{\columnwidth}{!}{\begin{tikzpicture}
\begin{axis}[
name=ax1,
width=.55\linewidth,
height=.55\linewidth,
grid style={dotted,anthrazit},
xmajorgrids,
yminorticks=true,
ymajorgrids,
legend columns=2,
title={\small(128,64)-Polar-CRC8},
title style={at={(0.5,0.93)},anchor=south},
legend cell align={left},
legend style={font=\small, at={(0,-0.33)},anchor=north west, /tikz/every even column/.append style={column sep=1.2cm}},
xlabel={\small$E_\mathrm{b}/N_0$ [dB]},
ylabel={\small Latency $\bar{\tau}$ [cycles]},
y label style={at={(axis description cs:-0.15,.5)},anchor=south},
ymode=log,
mark size=2pt,
xmin=1,
xmax=4,
ymin=50,
ymax=3000,
]

\addplot[color=mittelblau,line width = 1pt, dashed,mark=x, mark options={solid}]
table[col sep=comma, y expr=\thisrow{Y}*16+1]{
X, Y
1.00, 1.933e+02
1.50, 1.768e+02
2.00, 1.462e+02
2.50, 1.071e+02
3.00, 7.338e+01
3.50, 4.148e+01
4.00, 2.172e+01
};
\label{plot:128cabpl8lat}
\addlegendentry{\footnotesize CA-BPL-8, $\Pi$};

\addplot[color=mittelblau,line width = 1pt, solid,mark=x, mark options={solid}]
table[col sep=comma, y expr=\thisrow{Y}*16+1]{
X, Y
1.00, 1.869e+02
1.50, 1.668e+02
2.00, 1.481e+02
2.50, 1.107e+02
3.00, 7.007e+01
3.50, 4.018e+01
4.00, 2.070e+01
};
\label{plot:128cabpl8lat}
\addlegendentry{\footnotesize R-CA-BPL-8, $\Pi$};

\addplot[color=apfelgruen,line width = 1pt, dashed,mark=triangle, mark options={solid}]
table[col sep=comma, y expr=\thisrow{Y}*16+1]{
X, Y
1.00, 1.624e+02
1.50, 1.248e+02
2.00, 8.012e+01
2.50, 4.220e+01
3.00, 2.338e+01
3.50, 1.163e+01
4.00, 6.469e+00
};
\label{plot:128cabpl8lat}
\addlegendentry{\footnotesize P-CA-BPL-8, LTA};

\addplot[color=apfelgruen,line width = 1pt, solid,mark=triangle, mark options={solid}]
table[col sep=comma, y expr=\thisrow{Y}*16+1]{
X, Y
1.00, 1.594e+02
1.50, 1.184e+02
2.00, 7.242e+01
2.50, 3.528e+01
3.00, 1.889e+01
3.50, 9.205e+00
4.00, 5.441e+00
};
\label{plot:128cabpl8lat}
\addlegendentry{\footnotesize PR-CA-BPL-8, LTA};

\addplot[color=magenta,line width = 1pt, dotted,mark=none, mark options={solid}]
table[col sep=comma, y expr=\thisrow{Y}*16]{
X, Y
1.00, 1.354e+02
1.50, 8.691e+01
2.00, 4.523e+01
2.50, 2.168e+01
3.00, 1.034e+01
3.50, 5.506e+00
4.00, 3.604e+00
};
\label{plot:128cabplat}
\addlegendentry{\footnotesize R-CA-BP};

\addplot[color=rot,line width = 1pt, solid,mark=o, mark options={solid}]
table[col sep=comma]{
1.00, 320
1.50, 320
2.00, 320
2.50, 320
3.00, 320
3.50, 320
4.00, 320
};
\label{plot:128scl8}
\addlegendentry{\footnotesize CA-SCL-8};

\end{axis}


\begin{axis}[
title={\small(256,128)-Polar-CRC8},
title style={at={(0.5,0.93)},anchor=south},
at={(ax1.south east)},
xshift=0.8cm,
width=.55\linewidth,
height=.55\linewidth,
grid style={dotted,anthrazit},
xmajorgrids,
yminorticks=true,
ymajorgrids,
xlabel={$E_\mathrm{b}/N_0$ [dB]},
ymode=log,
mark size=2pt,
xmin=1,
xmax=4,
ymin=50,
ymax=3400,
]
\legend{};
\addplot[color=magenta,line width = 1pt, dotted,mark=none, mark options={solid}]
table[col sep=comma, y expr=\thisrow{Y}*18]{
X, Y
1.00, 1.405e+02
1.50, 8.092e+01
2.00, 3.235e+01
2.50, 1.259e+01
3.00, 6.230e+00
3.50, 4.060e+00
4.00, 3.450e+00
};
\label{plot:256cabplat}

\addplot[color=mittelblau,line width = 1pt, dashed,mark=x, mark options={solid}]
table[col sep=comma, y expr=\thisrow{Y}*18+1]{
X, Y
1.00, 1.977e+02
1.50, 1.922e+02
2.00, 1.779e+02
2.50, 1.483e+02
3.00, 9.829e+01
3.50, 5.655e+01
4.00, 3.122e+01
};
\label{plot:256cabpl8lat}

\addplot[color=mittelblau,line width = 1pt, solid,mark=x, mark options={solid}]
table[col sep=comma, y expr=\thisrow{Y}*18+1]{
X, Y
1.00, 1.969e+02
1.50, 1.929e+02
2.00, 1.784e+02
2.50, 1.452e+02
3.00, 1.103e+02
3.50, 6.411e+01
4.00, 3.310e+01
};
\label{plot:256cabpl8lat}

\addplot[color=apfelgruen,line width = 1pt, dashed,mark=triangle, mark options={solid}]
table[col sep=comma, y expr=\thisrow{Y}*18+1]{
X, Y
1.00, 1.689e+02
1.50, 1.213e+02
2.00, 5.560e+01
2.50, 2.724e+01
3.00, 1.215e+01
3.50, 6.787e+00
4.00, 4.496e+00
};
\label{plot:256cabpl8lat}

\addplot[color=apfelgruen,line width = 1pt, solid,mark=triangle, mark options={solid}]
table[col sep=comma, y expr=\thisrow{Y}*18+1]{
X, Y
1.00, 1.680e+02
1.50, 1.201e+02
2.00, 5.541e+01
2.50, 2.004e+01
3.00, 9.682e+00
3.50, 5.863e+00
4.00, 4.253e+00
};
\label{plot:256cabpl8lat}

\addplot[color=rot,line width = 1pt, solid,mark=o, mark options={solid}]
table[col sep=comma]{
1.00, 640
1.50, 640
2.00, 640
2.50, 640
3.00, 640
3.50, 640
4.00, 640
};
\label{plot:256scl8}

\end{axis}

\end{tikzpicture}}
	\caption{\footnotesize Decoding latency comparison of iterative and \ac{CA-SCL} decoding. All iterative decoders are limited to $N_\mathrm{it,max}=200$ iterations and use early stopping. As the blocklength $N$ increases, the latency gains due to the usage of iterative decoders become more pronounced.}
	\label{fig:latency}
\end{figure}
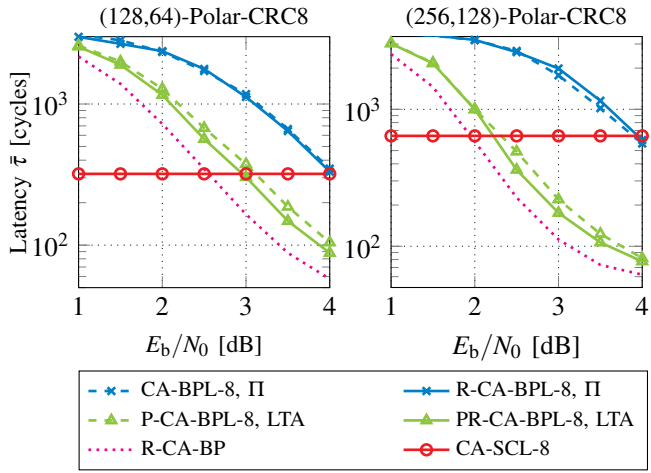
Next, we compare the decoding latency of the different iterative decoding schemes and \ac{CA-SCL} decoding. We assume a fully parallel implementation of each decoding algorithm, i.e., all operations that do not depend on each other can be executed at the same time and a single stage update takes one clock cycle. For iterative decoders, we assume a stopping criterion based on $\mathbf{\hat{v}}\mathbf{G}_N=\mathbf{\hat{c}}$ and the \ac{CRC} check is implemented.
The average latency of a CA-BP decoder is the average number of iterations $N_\mathrm{it,avg}$ times the latency of a single iteration, consisting of $n$ left-to-right updates, $n$ right-to-left updates and two cycles for the outer \ac{SPA} update. Hence, we have the average latency
\begin{equation}
    \bar{\tau}_{\mathrm{CA-BP}} = \operatorname{E}\left[(2n+2)\cdot N_\mathrm{it}\right] = (2n+2)\cdot N_\mathrm{it, avg}.
\end{equation}
For the ensemble variants (i.e., x-CA-BPL), we have to wait until all $L$ decoders have converged. Hence, we need to consider the average \textit{maximum} number of iterations. Additionally, we assume an extra cycle is required to select the final candidate. The latency is therefore
\begin{equation}
    \bar{\tau}_{\mathrm{CA-BPL}} = \operatorname{E}\left[(2n+2)\cdot \max_{i=1,\dots,L}\{N_{\mathrm{it},i}\}+1\right].\label{eq:lat_bpl}
\end{equation}
Lastly, for \ac{CA-SCL} decoding, there are $2N-1$ stage updates required, as well as $k$ additional cycles for metric sorting, and a final cycle for candidate selection \cite{Stimming2016HardwareIA}. Therefore, we have
\begin{equation}
    \bar{\tau}_{\mathrm{CA-SCL}} = 2N+k = (2+R)N.\label{eq:lat_scl}
\end{equation}
Fig. \ref{fig:latency} shows the latency of the compared decoders with respect to the \ac{SNR}. As expected, the single \ac{R-CA-BP} decoder has the lowest latency. Moreover, we can see that the proposed \ac{LTA} permuted decoders generally have a significantly lower latency than stage-shuffle permutations used in \ac{CA-BPL}. In contrast, relaxation only has a very minor impact on the convergence speed. The proposed \ac{PR-CA-BPL} algorithm outperforms \ac{CA-SCL} in terms of latency in the medium-to-high \ac{SNR} regime by a big margin. Additionally, as can be seen from Eq.~(\ref{eq:lat_bpl}), Eq.~(\ref{eq:lat_scl}) and Fig.~\ref{fig:latency}, the latency benefits of \ac{BPL}-based decoders over \ac{SCL} is larger for longer blocklengths $N$.

\section{Conclusion}\label{sec:conc}
In this work, we extend the \ac{CA-BPL} decoding algorithm to use \ac{LTA} permutations rather than stage-shuffles. Combined with factor graph relaxation, the proposed \ac{PR-CA-BPL} algorithm outperforms \ac{CA-BPL} both in error-rate performance and complexity. Overall, an error-rate performance similar to \ac{CA-SCL} can be reached, at a significantly lower decoding latency in the operating regime. Possible extensions of this work include the optimization of the used permutations (either for latency or error-rate performance) and the application of the proposed method to other pre-transformed polar codes, such as Polar-LDPC \cite{BP_Siegel_Concatenating} codes or \ac{PAC} codes, which can be described in the same polar subcode framework.
\bibliographystyle{IEEEtran}
\bibliography{references}
\end{NoHyper}
\end{document}